\documentclass[10pt,conference,A4]{IEEEtran}

\usepackage{amsmath}
\usepackage{url}
\usepackage[pdftex]{graphicx}
\usepackage[utf8x]{inputenc}
\usepackage[warn]{textcomp}% це для № (\textnumero), внизу
\usepackage{graphicx}
\usepackage{amsfonts}
\usepackage{bm}
\usepackage{statmath}

\begin{document}

\title{Harnessing thermal fluctuations for selectivity gain}

\author{\IEEEauthorblockN{Alexander Vidybida}
\IEEEauthorblockA{Bogolyubov Institute for Theoretical Physics
of the National Academy of Sciences of Ukraine,
Kyiv, Ukraine 03143\\
Email: vidybida@bitp.kiev.ua,vidybidaok@gmail.com, \url{http://vidybida.kiev.ua}}
}

\maketitle

\begin{abstract}

Selectivity of olfactory receptor neuron (ORN) is compared with
that of its receptor proteins (\textbf{\textit{R}}) with fluctuations 
of odor binding-releasing process taken into account.
The binding-releasing process is modeled as
\textbf{\textit{N}} Bernoulli trials, where \textbf{\textit{N}} is the total number of \textbf{\textit{R}}
per ORN. Dimensionless selectivities for both \textbf{\textit{R}} and ORN
are introduced and compared with each other.
It is found the ORN's selectivity can be much higher then that
of its receptor proteins. This effect is concentration-dependent.
Possible application for biosensors is discussed.
\end{abstract}

% no keywords

\section{Introduction}

In order to be useful, a chemical sensor must be able to discriminate
between different substances it is exposed to.
This ability, or selectivity can be achieved by different mechanisms.
In industrial biosensors, high selectivity is usually ensured at the first stage of
interaction between analyte and sensor. This is realized by covering with
highly selective molecules the primary sensing surface. Those molecules
could be highly specific enzymes \cite{Kucherenko2020}, or antibodies \cite{Cristea2015}.
%or other complex biological molecules suitable for a specific task, e.g., 
%virus detection, \cite{Zou2019}. 

For monitoring environment, another architecture has been developed in the olfactory
system. Olfactory receptor neuron (ORN) is considered in sensory biology as the 
primary reception unit. And typical ORN is rather generalist then specialist 
\cite{Buck2000,Su2009}. That means that an ORN increases its firing rate when exposed
to different odors. The degree of increase depends on the substance presented,
and this determines ORN's selectivity. 
Normally, discriminating ability increases when olfactory signal travels from
primary reception units to higher brain areas, see Table \ref{Tab1}.
It is well known that selectivity of a projection neuron, which receives stimulation directly 
from ORNs,
is better than that of ORNs converging on it, \cite{Duchamp-Viret1990}. At high odor concentration
this happens due to mechanism of lateral inhibition in the olfactory bulb \cite{Linster2005,Valley2008}.
At low odor concentration, when lateral inhibition seems not working, \cite{Duchamp1982}, another interesting mechanism has been proposed in \cite{Vidybida2019a}.
Further, in the olfactory cortex, each scent is represented by specific activity in a neuronal assembly,
and each neuron is involved in representation of many odors, \cite{Malnic1999}.
Spatio-temporal pattern of activity in the assembly is essential for this final odor recognition
\cite{Stopfer1997,Laurent2002,Wilson2011}.

From the physical point of view, the primary perception of odors happens in the set of $N$ identical
receptor proteins, $R$, expressed in the ORN's cilia. This set, or any individual $R$ from it
has its own selectivity, which is based on different chemical affinity between different odors
and $R$. Is this selectivity the same as that of the corresponding ORN? Comparison of the two
selectivities is not a trivial task due to different physical nature of response, 
see Table \ref{Tab1}.
But any quantitative measure of selectivity can/should be expressed in dimensionless units.
This allows for comparing selectivities for systems with qualitatively different physical nature 
of response as in the case of $R$ and ORN. Below, we define such a dimensionless measure of selectivity for both $R$ and
ORN, and compare them. In this course we take into account that binding-releasing of odor
molecules with $R$ is subjected to thermal fluctuations since it is driven by Brownian motion.
Also, we take into account that ORN is highly nonlinear processing unit due to 
presence of the firing threshold. Finally, we prove that due to this features the selectivity 
of ORN can be considerably higher than that of its receptor proteins. 
The degree of selectivity gain is calculated exactly for a simple ORN model.
Possible implementation of this effect
in artificial biosensors is discussed in conclusion.
{\small 
\begin{table}
\center
\begin{tabular}{c | c}
{\bf constructive element} & {\bf measure of response} \\
\hline
&\\
receptor proteins & fraction of bound receptors\\
$\displaystyle \downarrow$ &                                   \\
receptor neurons & {mean firing rate}\\
$\displaystyle \downarrow$ &                                   \\
projection neurons & {mean firing rate}\\
$\displaystyle \downarrow$ &                                   \\
olfactory cortex & activity in local cortical\\
        & circuits (combinatorial code)\\
\hline
\end{tabular}\medskip
\caption{\label{Tab1}Selectivity build up steps in a biological olfactory system}
\end{table} 
}

\section{Definitions and assumptions}
\subsection{Model of ORN}\label{ORN}
In order to make possible a simple
exact mathematical analysis we use an extremely simplified model.
Namely, in this model, ORN has $N$ identical receptor proteins $R$
able to bind reversibly with odor molecules $A$. 
Each bound $R$ contributes the same amount to the receptor potential, which is
depolarization of ORN's excitable membrane. 
If the number $n(t)$ of bound $R$ is above or equal to $N_0$, where $N_0<N$, 
the firing threshold is achieved
and the ORN fires output spikes with a constant frequency $f$. 
Otherwise, it is silent. We assume that binding-releasing of odor at any 
individual $R$
is statistically independent of what happens with other receptor proteins.

\subsection{Selectivity of receptor proteins}
Consider two separate experiments in which two odors $A_1$ and $A_2$ are presented 
at the same concentration $c$ to the set of $R$. Due to thermal fluctuations, 
the instantaneous number $n(t)$ of bound $R$ will change randomly. 
In the equilibrium, the mean number of bound receptors is $p_iN$, where
$0\le p_i\le 1$, $i=1,2$. Any of $p_i$ can be found as $p=c/(c+K_D)$,
where $K_D$ is the corresponding dissociation constant.
If $A_1$ has more affinity with $R$ than does $A_2$, then $p_1>p_2$:
\begin{equation}\label{ineq}
p_1=p_2+\Delta p, \quad \Delta p >0.
\end{equation}
In this case we say that $R$ is able to discriminate between $A_1$ and $A_2$
and characterize this ability by the following dimensionless selectivity:
\begin{equation}\label{SR}
%S_R=\frac{\Delta p}{p_1}.
S_R=\Delta p\,/\,p_1.
\end{equation}
Actually, the quantity $p$ here gives the probability that any $R$ is bound with odor
molecule (binding probability) if observed at any moment.

\subsection{Selectivity of ORN}
We assume here that the concentration $c$ ensures that mean number of bound 
receptors $p_1N$, $p_2N$ is close to the firing threshold $N_0$. In this case, the instantaneous
number $n(t)$ will cross the threshold $N_0$ randomly due to thermal fluctuations 
both for $A_1$ and $A_2$.
If we observe the ORN activity during some fixed time interval $T$, its mean firing rate
will be $F=fT_a/T$, where $T_a<T$ is the total time the $n(t)$ spends above $N_0$,
both for $A_1$ and $A_2$. 
From (\ref{ineq}) it follows that $T_{a1}>T_{a2}$, which results in $F_1>F_2$.
The latter means that ORN is able to discriminate between $A_1$ and $A_2$.
The dimensionless selectivity of ORN is as follows
\begin{equation}\label{SORN}
%S_{ORN}=\frac{F_1-F_2}{F_1}=\frac{\Delta T_a}{T_{a1}},
S_{ORN}=(F_1-F_2)\,/\,F_1=\Delta T_a\,/\,T_{a1},
\end{equation}
where $\Delta T_a = T_{a1}-T_{a2}$.

\section{Selectivity gain}
Here we compare selectivity of ORN with that of its
receptor proteins. For this purpose define selectivity gain $g$
as follows:
\begin{equation}\label{g}
%g=\frac{S_{ORN}}{S_R} = \frac{\Delta T_a\, p_1}{\Delta p\, T_{a1}}.
g=S_{ORN}\,/\,S_R = (\Delta T_a\,p_1)\,/\,(\Delta p\,T_{a1}).
\end{equation}
For {\it poor selectivities} both $\Delta p$ and $\Delta T_a$ are small.
 Taking this into account the latter can be rewritten as a derivative:
 \begin{equation}\label{g2}
 g(p)=\frac{p}{T_a}\frac{d\,T_a}{d\,p},
 \end{equation}
 where $T_a$ is the amount of time spent above the threshold
 during period $T$ for a given  binding probability $p$.

 It seems evident that
 \begin{equation}\label{TaProb}
 T_a=T\,\mathbf{Prob}\{n(t)\ge N_0\},
 \end{equation}
 where
 \begin{equation}\label{Prob}
 \mathbf{Prob}\{n(t)\ge N_0\}=
 \sum\limits_{N_0\le k\le N}\binom{N}{k}p^k(1-p)^{N-k}.
 \end{equation}
 Eq. (\ref{g2}), after substituting (\ref{TaProb}) and (\ref{Prob})
 turns into the following:
{\small\begin{equation}\label{gp}
  g(p)= 
  \frac{p
\sum\limits_{N_0\le k\le N} \frac{1}{k!(N-k)!}p^{k-1}(1-p)^{N-k-1}(k-Np)
  }{
  \sum\limits_{N_0\le k\le N} \frac{1}{k!(N-k)!}p^k(1-p)^{N-k}
  }\,.
\end{equation}}
A more compact formula can be obtained
from (\ref{gp}) by using binomial cumulative probability function:
$$
cdf(N_0,N,p)=\sum\limits_{0\le k\le N_0}\binom{N}{k} p^k(1-p)^{N-k}.
$$
By applying this in (\ref{gp}) one obtains after transformations:
\begin{equation}\label{gbccnq}
  g(p)=   
  N\frac{p}{1-p}\left(\frac
     {cdf(N-N0, N - 1, 1-p)}
     {cdf(N-N0, N, 1-p)}
     -1\right)\,.
\end{equation}

 \section{Numerical examples}
 
\begin{figure}
\center
\includegraphics[width=0.3\textwidth]{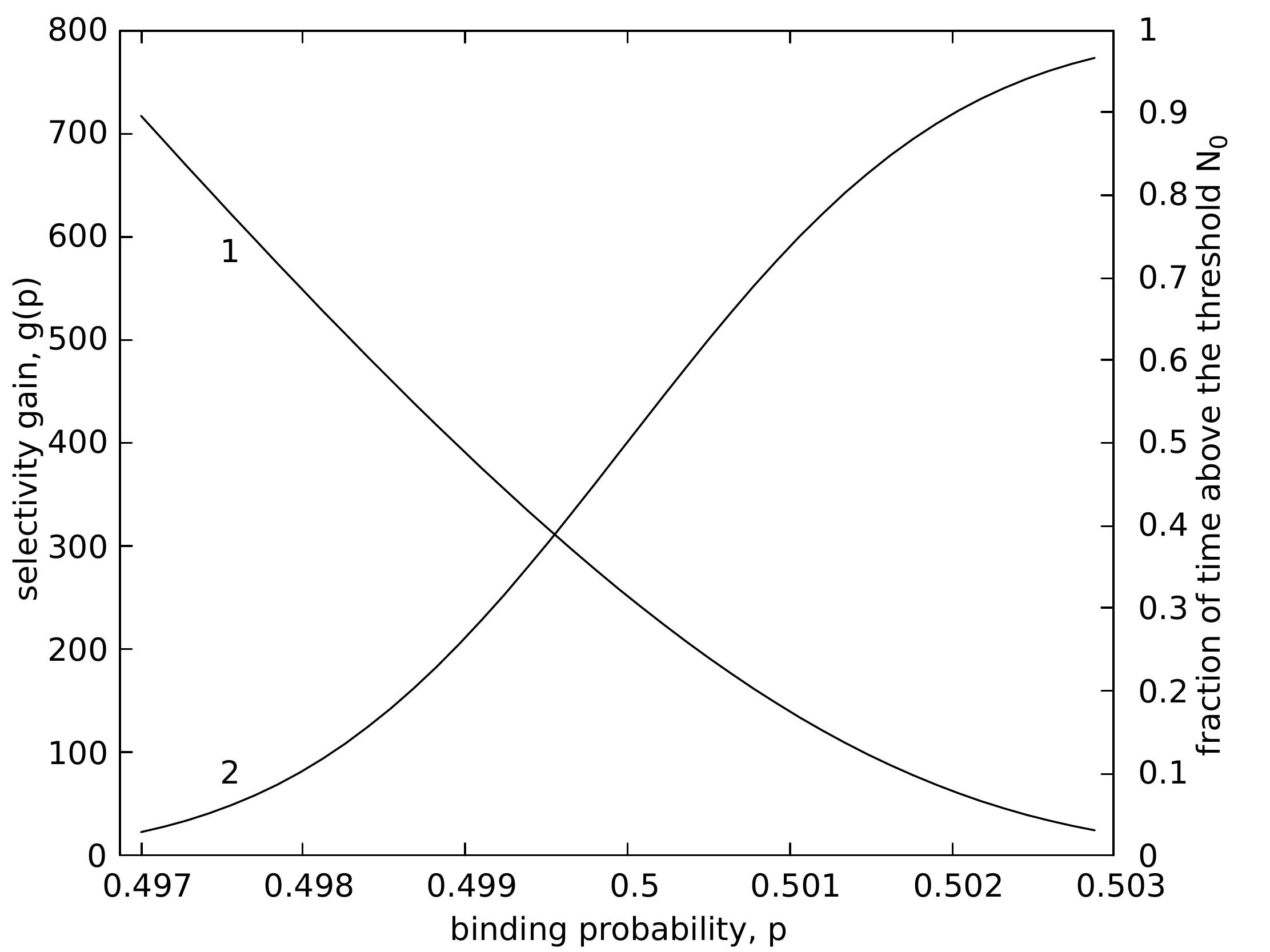}
\caption{\label{Fig1}Dependence of selectivity gain, 1, and fraction of time spent above the
threshold, 2, on the binding probability. Here $N = 100\,000$,  $N_0 = 50\,000$.}
\end{figure}

 Numerical estimates of the derivative $dg(p)/dp$ for several sets of pares $(N,N_0)$
 supports the idea that
 $$
dg(p)\,/\,dp < 0,\quad 0\le p\le 1.
 $$
Since $p$ increases with odor concentration, from the latter, one could expect
higher selectivity gain for smaller concentrations. 
On the other hand, denominator $cdf(N-N0, N, 1-p)$ in (\ref{gbccnq}) gives exactly the fraction of time
$n(t)$ spends above the threshold. This fraction determines the ORN's level
of output (mean firing rate),
and it decreases with decreasing concentration and $p$.
 So, we have here a trade-off between selectivity
and sensitivity: with decreasing concentration and binding probability $p$
one gets higher selectivity, but lower sensitivity, see also Fig \ref{Fig1}. 
Which values of $p$
might be of practical interest depends on concrete values of the total number
$N$ of binding sites in a sensor and the minimal number $N_0$ of bound sites
required for having activity at its output end.

In the Table \ref{Tab2} we put some 
examples
%\footnote{The values are calculated using {\tt cdf\_binomial(x,n,p)}
%function in the computer algebra system {\tt maxima} with {\tt fpprec} set to 100.}
 of selectivity gain.
The first two rows of the Table \ref{Tab2} give examples for a moth
pheromone ORN. 
The possible value for the number of receptor proteins
$R$ per ORN is taken from \cite{Kaissling2001}. 
Based on data from \cite{Kaissling2001} it is possible to 
%conclude\footnote{J.-P. Rospars, private communication.}
conclude
%\footnote{J.-P. Rospars, private communication.}
that several hundreds could be a good value for $N_0$. Actually, the threshold 
value for a biological ORN depends on many factors not considered here, like ionic
composition on both sides of the excitable membrane,
which may be variable, and presence of inhibitory stimulation,
which can be fast or slow.
Therefore, values in Table \ref{Tab2} are rather illustrative than conclusive.
{\small
\begin{table}
\center
\begin{tabular}{rrccc}
$N$ & $N_0$ & $p$ & $g(p)$ & time above $N_0$ \\
\hline
2556000 & 200 & 7.5$\cdot10^{-5}$& 16.6 & 28\%\\
2556000 & 250 & 7.5$\cdot10^{-5}$& 61 & 0.03\%\\
100000 & 10000 & 0.1 & 84 & 50\%\\
1000000 & 100000 & 0.1 & 266 & 50\%\\
10000000 & 1000000 & 0.1 & 841 & 50\%\\
\hline
\end{tabular}\medskip
\caption{\label{Tab2}Selectivity gain examples}
\end{table} 
}
 \section{Conclusions and discussion}
In this note, we discussed a possible utilization of thermal noise for improving
odor discrimination. Consideration is made for a simplified model of olfactory receptor neuron.
Selectivity of ORN is compared with that of its receptor proteins. It is concluded that 
the former can be much higher than the latter if thermal fluctuations 
of odor molecules binding-releasing process are taken into account.

This is not the only case when Brownian motion is used beneficially in living
objects, see e.g. \cite{Yanagida2017} for muscles contraction. 
Also, this is not the only case when fluctuations processing instead of filtering them out 
is proposed for improving discriminating ability. Actually, this note falls into the
promising and developed area of fluctuation enhanced sensing, see \cite{SMULKO2001,Scandurra2020}.

The model of ORN described in Sec. \ref{ORN} is a very simplified toy model.
Actually, under receptor protein $R$ we have in mind a ligand-gated ion channel
 found in insects, \cite{Sato2008}. We leave in the model the main
source of non-linearity, namely, the firing threshold and this is enough
to demonstrate the idea. Another sources of non-linearity can be found in
ORN due to its internal biochemical mechanisms, especially, 
if ORN expresses G protein–coupled receptors, \cite{Munger2012}.
This additional non-linearity seems working in the same direction as the main one:
improving ORN selectivity as compared with that of its receptor proteins.

A less evident limitation is that the effect of selectivity gain 
considered here can be observed
in the narrow range of odor concentrations, or binding probabilities $p$, see 
Fig. \ref{Fig1}. This range depends on the firing threshold level $N_0$.
The latter is variable in living objects due to adaptation and inhibition
\cite{McGann2013}. In practical realizations, intended to work in wide concentration range,
 a possibility of tuneable
threshold should be considered. 
Actually, a scent description includes both identity (odor species) and intensity
(concentrations). This note offers nothing as regards concentration. Understanding
how odor intensity is represented in olfactory system might help to resolve the above
mentioned limitation. Possible steps in this direction are made in \cite{Lnsk1993,Bolding2017}.

Size of industrial biosensors constantly decreases.
In small devices, noise represents an essential part of output.
Utilizing noise for improving device characteristics could be a better choice
than averaging out signal fluctuations.
As we can see from three bottom rows of Table  \ref{Tab2}, selectivity gain due to the
mechanism discussed here can be quite large. 
Exploiting this mechanism opens a perspective of using as primary
odor recognition sites in biosensors
simple molecules with low discriminating ability, but cheaper,
and  with better performance characteristics like durability and robustness.
High selectivity can be achieved due to selectivity gain 
at further stages of noisy signal processing.
\smallskip

{\small \linespread{1.6}
\textit{Acknowledgments}. This work was supported by the Programs of Basic Research of the Department of Physics and Astronomy of the National Academy of Sciences of Ukraine ``Mathematical models of nonequilibrium processes in open systems'', № 0120U100857, and ``Noise-induced dynamics and correlations in nonequilibrium systems'', № 0120U101347.
}


\begin{thebibliography}{10}

\bibitem{Kucherenko2020}
I.~S. Kucherenko, O.~O. Soldatkin, S.~V. Dzyadevych, and A.~P. Soldatkin,
  ``Electrochemical biosensors based on multienzyme systems: Main groups,
  advantages and limitations – a review,'' {\em Analytica Chimica Acta},
  vol.~1111, pp.~114--131, 2020.

\bibitem{Cristea2015}
C.~Cristea, A.~Florea, M.~Tertiș, and R.~Sandulescu,
``Ch. 6 - Immunosensors,'' in {\em Biosensors}
  (T.~Rinken, ed.),  IntechOpen, 2015.

\bibitem{Buck2000}
L.~B. Buck, ``The molecular architecture of odor and pheromone sensing in
  mammals,'' {\em Cell}, vol.~100, pp.~611--618, 2000.

\bibitem{Su2009}
C.-Y. Su, K.~Menuz, and J.~R. Carlson, ``Olfactory perception: Receptors,
  cells, and circuits,'' {\em Cell}, vol.~139, no.~1, pp.~45--59, 2009.

\bibitem{Duchamp-Viret1990}
P.~Duchamp-Viret, A.~Duchamp, and G.~Sicard, ``Olfactory discrimination over a
  wide concentration range. comparison of receptor cell and bulb neuron
  abilities,'' {\em Brain Research}, vol.~517, pp.~256--262, 1990.

\bibitem{Linster2005}
T.~A. Cleland and C.~Linster, ``Computation in the olfactory system,'' {\em
  Chemical Senses}, vol.~30, no.~9, pp.~801--813, 2005.

\bibitem{Valley2008}
M.~T. Valley and S.~Firestein, ``A lateral look at olfactory bulb lateral
  inhibition,'' {\em Neuron}, vol.~59, no.~5, pp.~682--684, 2008.

\bibitem{Duchamp1982}
A.~Duchamp, ``Electrophysiological responses of olfactory bulb neurons to odour
  stimuli in the frog. a comparison with receptor cells,'' {\em Chemical
  Senses}, vol.~7, no.~2, pp.~191--210, 1982.

\bibitem{Vidybida2019a}
A.~K. Vidybida, ``Possible stochastic mechanism for improving the selectivity
  of olfactory projection neurons,'' {\em Neurophysiology}, vol.~51, no.~3,
  pp.~152--159, 2019.

\bibitem{Malnic1999}
B.~Malnic, J.~Hirono, T.~Sato, and L.~B. Buck, ``Combinatorial receptor codes
  for odors,'' {\em Cell}, vol.~96, p.~713 723, 1999.

\bibitem{Stopfer1997}
M.~Stopfer, S.~Bhagavan, B.~H. Smith, and G.~Laurent, ``Impaired odour
  discrimination on desynchronization of odour-encoding neural assemblies,''
  {\em Nature}, vol.~390, pp.~70--74, 1997.

\bibitem{Laurent2002}
G.~Laurent, ``Olfactory network dynamics and the coding of multidimensional
  signals,'' {\em Nature Reviews Neuroscience}, vol.~3, no.~11, pp.~884--895,
  2002.

\bibitem{Wilson2011}
D.~A. Wilson and R.~M. Sullivan, ``Cortical processing of odor objects,'' {\em
  Neuron}, vol.~72, no.~4, pp.~506--519, 2011.

\bibitem{Kaissling2001}
K.-E. Kaissling, ``Olfactory perireceptor and receptor events in moths: A
  kinetic model,'' {\em Chemical Senses}, vol.~26, pp.~125--150, 2001.

\bibitem{Yanagida2017}
T.~Yanagida and Y.~Ishii, ``Single molecule detection, thermal fluctuation and
  life,'' {\em Proceedings of the Japan Academy. Series B, Physical and
  biological sciences}, vol.~93, no.~2, pp.~51--63, 2017.

\bibitem{SMULKO2001}
J.~Smulko, C.-G. Granqvist, and L.~B. Kish, ``On the statistical analysis of
  noise in chemical sensors and its application for sensing,'' {\em Fluctuation
  and Noise Letters}, vol.~01, no.~03, pp.~L147--L153, 2001.

\bibitem{Scandurra2020}
G.~Scandurra, J.~Smulko, and L.~B. Kish, ``Fluctuation-enhanced sensing (FES):
  A promising sensing technique,'' {\em Applied Sciences}, vol.~10, no.~17,
  2020.

\bibitem{Sato2008}
K.~Sato, M.~Pellegrino, T.~Nakagawa, T.~Nakagawa, L.~B. Vosshall, and
  K.~Touhara, ``Insect olfactory receptors are heteromeric ligand-gated ion
  channels,'' {\em Nature}, vol.~452, no.~7190, pp.~1002--1006, 2008.

\bibitem{Munger2012}
S.~D. Munger, ``Chapter 52 - molecular basis of olfaction and taste,'' in {\em
  Basic Neurochemistry (Eighth Edition)} (S.~T. Brady, G.~J. Siegel, R.~W.
  Albers, and D.~L. Price, eds.), pp.~904--915, Academic Press, 2012.

\bibitem{McGann2013}
J.~P. McGann, ``Presynaptic inhibition of olfactory sensory neurons: new
  mechanisms and potential functions,'' {\em Chemical Senses}, vol.~38, no.~6,
  pp.~459--474, 2013.

\bibitem{Lnsk1993}
P.~Lánský and J.-P. Rospars, ``Coding of odor intensity,'' {\em BioSystems},
  vol.~31, pp.~15--38, 1993.

\bibitem{Bolding2017}
K.~A. Bolding and K.~M. Franks, ``Complementary codes for odor identity and
  intensity in olfactory cortex,'' {\em eLife}, vol.~6, p.~e22630, 2017.

\end{thebibliography}
\end{document}